\documentclass[a4paper, 11pt]{article}

\usepackage[utf8x]{inputenc}
\usepackage{epsfig,amssymb,euscript,xspace, color}
\usepackage{jheppub}
\usepackage[T1]{fontenc}
\usepackage{color}
\usepackage{amssymb}
\usepackage{amsfonts,amsmath,bm}
\usepackage{graphicx}
\usepackage{mathtools}
\usepackage{ragged2e}

\usepackage{dcolumn}
\usepackage{graphicx}
\usepackage{braket}
\usepackage{verbatim}
\usepackage[english]{babel}
\usepackage{hyperref}
\hypersetup{
citecolor=red,
colorlinks=true,
filecolor=red,
linkcolor=blue,
linktocpage=true,
urlcolor=blue
}
\usepackage[titletoc,toc,title]{appendix}
\numberwithin{equation}{section}
\usepackage{cleveref}
\usepackage{epsfig}
\usepackage{float}
\usepackage{amsfonts}
\usepackage{enumitem}
\usepackage[font={footnotesize,it}]{caption}
\usepackage{cases}

\newcommand{\be}{\begin{equation}}
\newcommand{\ee}{\end{equation}}
\newcommand{\eq}[1]{(\ref{#1})}
\newcommand\trick[1]{}
\newcommand{\bit}{\begin{itemize}}  \newcommand{\eit}{\end{itemize}}
\newcommand{\ben}{\begin{enumerate}}  \newcommand{\een}{\end{enumerate}}

\newcommand{\rf}[1]{(\ref{#1})}

\def\bd{\begin{document}}
\def\ed{\end{document}}
\def\bea{\begin{eqnarray}}
\def\eea{\end{eqnarray}}
\let\bm=\bibitem

\def\la{\langle}
\def\ra{\rangle}

\def\npb#1#2#3{Nucl. Phys. {\bf{B#1}} #3 (#2)}
\def\plb#1#2#3{Phys. Lett. {\bf{#1B}} #3 (#2)}
\def\prl#1#2#3{Phys. Rev. Lett. {\bf{#1}} #3 (#2)}
\def\prd#1#2#3{Phys. Rev. {D bf{#1}} #3 (#2)}
\def\cmp#1#2#3{Comm. Math. Phys. {\bf{#1}} #3 (#2)}
\def\cqg#1#2#3{Class. Quantum Grav. {\bf{#1}} #3 (#2)}
\def\nppsa#1#2#3{Nucl. Phys. B (Proc. Suppl.) {\bf{#1A}}#3 (#2)}
\def\ap#1#2#3{Ann. of Phys. {\bf{#1}} #3 (#2)}
\def\ijmp#1#2#3{Int. J. Mod. Phys. {\bf{A#1}} #3 (#2)}
\def\rmp#1#2#3{Rev. Mod. Phys. {\bf{#1}} #3 (#2)}
\def\mpla#1#2#3{Mod. Phys. Lett. {\bf A#1} #3 (#2)}
\def\jhep#1#2#3{J. High Energy Phys. {\bf #1} #3 (#2)}
\def\atmp#1#2#3{Adv. Theor. Math. Phys. {\bf #1} #3 (#2)}

\def\sst{\scriptscriptstyle}
\def\thetabar{\bar\theta}
\def\Tr{{\rm Tr}}
\def\one{\mbox{1 \kern-.59em {\rm l}}}

\def\a{\alpha}      \def\da{{\dot\alpha}}  \def\dA{{\dot A}}
\def\b{\beta}       \def\db{{\dot\beta}}
\def\g{\gamma}  \def\G{\Gamma}  \def\dc{{\dot\gamma}}
\def\d{\delta}  \def\D{\Delta}  \def\ddt{\dot\delta}
\def\e{\epsilon}
\def\ve{\varepsilon}
\def\uve{\upvarepsilon}
\def\f{\phi}    \def\F{\Phi}    \def\vvf{\f}
\def\vphi{\varphi}
\def\h{\eta}
\def\k{\kappa}
\def\l{\lambda} \def\L{\Lambda}
\def\m{\mu} \def\n{\nu}
\def\o{\omega}
\def\p{\pi} \def\P{\Pi}
\def\r{\rho}
\def\s{\sigma}  \def\S{\Sigma}
\def\t{\tau}
\def\th{\theta} \def\Th{\Theta} \def\vth{\vartheta}
\def\X{\Xeta}
\def\z{\zeta}

\def\na{\nabla}

\def\cA{{\cal A}} \def\cB{{\cal B}} \def\cC{{\cal C}}
\def\cD{{\cal D}} \def\cE{{\cal E}} \def\cF{{\cal F}}
\def\cG{{\cal G}} \def\cH{{\cal H}} \def\cI{{\cal I}}
\def\cJ{{\mathscr J}} \def\cK{{\cal K}} \def\cL{{\cal L}}
\def\cM{{\cal M}} \def\cN{{\cal N}} \def\cO{{\cal O}}
\def\cP{{\cal P}} \def\cQ{{\cal Q}} \def\cR{{\cal R}}
\def\cS{{\cal S}} \def\cT{{\cal T}} \def\cU{{\cal U}}
\def\cV{{\cal V}} \def\cW{{\cal W}} \def\cX{{\cal X}}
\def\cY{{\cal Y}} \def\cZ{{\cal Z}}
\def\ct{{\cal t}}

\def\ua{\underline{\alpha}}
\def\uc{\underline{\phantom{\alpha}}\!\!\!\gamma}
\def\um{\underline{\mu}}
\def\ud{\underline\delta}
\def\ue{\underline\epsilon}
\def\una{\underline a}\def\unA{\underline A}
\def\unb{\underline b}\def\unB{\underline B}
\def\unc{\underline c}\def\unC{\underline C}
\def\und{\underline d}\def\unD{\underline D}
\def\une{\underline e}\def\unE{\underline E}
\def\unf{\underline{\phantom{e}}\!\!\!\! f}\def\unF{\underline F}
\def\unm{\underline m}\def\unM{{\underline M}}
\def\unn{\underline n}\def\unN{{\underline N}}
\def\unp{\underline{\phantom{a}}\!\!\! p}\def\unP{\underline P}
\def\unq{\underline{\phantom{a}}\!\!\! q}
\def\unQ{\underline{\phantom{A}}\!\!\!\! Q}
\def\unH{\underline{H}}

\def\As {{A \hspace{-6.4pt} \slash}\;}
\def\bs {{b \hspace{-6.4pt} \slash}\;}
\def\Ds {{D \hspace{-6.4pt} \slash}\;}
\def\Gts {{\Gt \hspace{-6.4pt} \slash}\;}
\def\ds {{\del \hspace{-6.4pt} \slash}\;}
\def\ss {{\s \hspace{-6.4pt} \slash}\;}
\def\ks {{ k \hspace{-6.4pt} \slash}\;}
\def\ps {{p \hspace{-6.4pt} \slash}\;}
\def\xs {{x \hspace{-6.4pt} \slash}\;}
\def\pas {{{p_1} \hspace{-6.4pt} \slash}\;}
\def\pbs {{{p_2} \hspace{-6.4pt} \slash}\;}
\def\cFs {{{\cal F} \hspace{-6.4pt} \slash}\;}
\def\Dss {{D \hspace{-7.5pt} \slash}\;}
\def\dss {{\del \hspace{-7.0pt} \slash}\;}

\def\Ah{{\hat{A}}}
\def\Dh{{\hat{D}}}
\def\Gh{{\hat{G}}}
\def\Fh{{\hat{F}}}
\def\Ih{{\hat{I}}}
\def\Jh{{\hat{J}}}
\def\Kh{{\hat{K}}}
\def\Lh{{\hat{L}}}
\def\Ph{{\hat{P}}}
\def\Rh{{\hat{R}}}
\def\Vh{{\hat{V}}}
\def\Xh{{\hat{X}}}

\def\ah{{\hat{\a}}}
\def\bh{{\hat{\b}}}
\def\gh{{\hat{\g}}}
\def\dh{{\hat{\d}}}
\def\rh{{\hat{\r}}}
\def\hh{\hat{h}}
\def\uh{\hat{u}}
\def\xh{\hat{x}}
\def\yh{\hat{y}}
\def\ph{\hat{p}}
\def\xih{\hat{\xi}}
\def\chih{\hat{\chi}}
\def\Psih{\hat{\Psi}}
\def\phih{\hat{\phi}}

\def\psit{\tilde{\psi}}
\def\Psit{\tilde{\Psi}}
\def\Psibt{\tilde{\bar{Psi}}}

\def\lambdat{\tilde {\lambda}}
\def\st{\tilde{\sigma}}

\def\delt{\tilde{\delta}}
\def\Phit{\tilde{\Phi}}
\def\Phitb{\overline{\tilde{Phi}}}
\def\tht{\tilde{\th}}
\def\lt{\tilde{\l}}
\def\chit{\tilde{\chi}}
\def\phit{\tilde{\phi}}

\def\At{\tilde{A}}
\def\Bt{\tilde{B}}
\def\Ct{\tilde{C}}
\def\Dt{\tilde{D}}
\def\Et{\tilde{E}}
\def\Ft{\tilde{F}}
\def\Gt{\tilde{G}}
\def\Ht{\tilde{H}}
\def\It{\tilde{I}}
\def\Jt{\tilde{J}}
\def\Pt{\tilde{P}}
\def\Ot{\tilde{O}}
\def\Mt{\tilde{M }}
\def\Nt{\tilde{N}}
\def\St{\tilde{S}}
\def\Vt{\tilde{V}}
\def\Xt{\tilde{X}}
\def\at{\tilde{a}}
\def\dt{\tilde{d}}
\def\htt{\tilde{h}}
\def\ft{\tilde{f}}
\def\gt{\tilde{g}}
\def\pt{\tilde{p}}
\def\qt{\tilde{q}}
\def\rt{\tilde{r}}
\def\nt{\tilde{n}}
\def\ut{\tilde{u}}
\def\wt{\tilde{w}}
\def\zt{\tilde{z}}
\def\xt{\tilde{x}}
\def\yt{\tilde{y}}
\def\Psit{\tilde{\Psi}}
\def\phit{\tilde{\phi}}
\def\tD{\tilde{\D}}

\def\eb{\bar{\epsilon}}
\def\delb{\bar{\partial}}
\def\thb{\bar{\theta}}
\def\mub{\bar{\mu}}
\def\lamb{\bar{\l}}
\def\psib{\bar{\psi}}
\def\sb{\bar{\sigma}}
\def\xib{\bar{\xi}}
\def\chib{\bar{\chi}}

\def\Psib{\bar{\Psi}}
\def\Phib{\bar{\Phi}}
\def\Lamb{\bar{\Lambda}}
\def\Sb{{\overline \Sigma}}
\def\cb{\bar{c}}
\def\hb{\bar{h}}
\def\qb{\bar{q}}
\def\wb{\bar{w}}
\def\ub{\bar{u}}
\def\zb{{\bar{z}}}
\def\Hb{\bar{H}}
\def\Qb{{\bar Q}}
\def\Omegab{\overline{\Omega}}
\def\ob{\overline{\omega}}

\def\Ab{{\overline A}} \def\Bb{{\overline B}} \def\Cb{{\overline C}}
\def\Db{{\overline D}} \def\Eb{{\overline E}} \def\Fb{{\overline F}}
\def\Gb{{\overline G}}
\def\Ib{{\overline I}}
\def\Jb{{\overline J}} \def\Kb{{\overline K}} \def\Lb{{\overline L}}
\def\Mb{{\overline M}} \def\Nb{{\overline N}} \def\Ob{{\overline O}}
\def\Pb{{\overline P}}  \def\Rb{{\overline R}}
 \def\Tb{{\overline T}} \def\Ub{{\overline U}}
\def\Vb{{\overline V}} \def\Wb{{\overline W}} \def\Xb{{\overline X}}
\def\Yb{{\overline Y}} \def\Zb{{\overline Z}}

\def\fb{{\overline f}}
\def\gb{{\overline g}}
\def\nb{{\overline n}}
\def\mb{{\overline m}}
\def\lb{{\overline l}}
\def\yb{{\overline y}}

\def\ldel{{\overleftarrow{\del}}}
\def\rdel{{\overrightarrow{\del}}}
\def\ldeldel{{\overleftarrow{\del^2}}}
\def\rdeldel{{\overrightarrow{\del^2}}}
\def\ldelb{{\overleftarrow{\bar{\del}}}}
\def\rdelb{{\overrightarrow{\bar{\del}}}}

\def\ba{{\bf a}}
\def\bk{{\bf k}}
\def\bl{{\bf l}}
\def\bp{{\bf p}}
\def\bq{{\bf q}}
\def\br{{\bf r}}
\def\bt{{\bf t}}
\def\bu{{\bf u}}
\def\bv{{\bf v}}
\def\bx{{\bf x}}
\def\by{{\bf y}}
\def\bA{{\bf A}}
\def\bR{{\bf R}}
\def\bV{{\bf V}}

\def\bz{{\boldsymbol{\zeta}}}

\def\bone{{\bf 1}}

\def\va{{\vec a}}
\def\vk{{\vec k}}
\def\vp{{\vec p}}
\def\vq{{\vec q}}
\def\vx{{\vec x}}
\def\vy{{\vec y}}
\def\vu{{\vec u}}
\def\vv{{\vec v}}
\def \vH{{\vec H}}
\def \vg{{\vec g}}

\def\vs{{\vec \sigma}}
\def\vtau{{\vec \tau}}

\newcommand{\ov}[1]{\overrightarrow{#1}}

\def\frA{\mathfrak{A}}
\def\frB{\mathfrak{B}}
\def\frC{\mathfrak{C}}
\def\frD{\mathfrak{D}}
\def\frE{\mathfrak{E}}
\def\frF{\mathfrak{F}}
\def\frG{\mathfrak{G}}
\def\frH{\mathfrak{H}}
\def\frM{\mathfrak{M}}
\def\frN{\mathfrak{N}}
\def\frR{\mathfrak{R}}
\def\frW{\mathfrak{W}}

\def\fra{\mathfrak{a}}
\def\frb{\mathfrak{b}}
\def\frf{\mathfrak{f}}
\def\frg{\mathfrak{g}}
\def\frh{\mathfrak{h}}
\def\frl{\mathfrak{l}}
\def\frs{\mathfrak{s}}
\def\fri{\mathfrak{i}}
\def\frj{\mathfrak{j}}

\def\ma{\mathfrak{a}}
\def\mg{\mathfrak{g}}
\def\mh{\mathfrak{h}}
\def\mR{\mathfrak{R}}
\def\mN{\mathfrak{N}}

\newcommand{\nn}{{\nonumber}}

\def\d{\delta}\def\D{\Delta}\def\ddt{\dot\delta}

\def\pa{\partial} \def\del{\partial}
\def\xx{\times}
\def\uno{\mbox{1 \kern-.59em {\rm l}}}

\def\trp{^{\top}}
\def\inv{^{-1}}
\def\dag{\dagger}
\def\pr{^{\prime}}

\def\rar{\rightarrow}
\def\lar{\leftarrow}
\def\lrar{\leftrightarrow}

\newcommand{\0}{\,\!}      
\def\one{1\!\!1\,\,}
\def\im{\imath}
\def\jm{\jmath}

\newcommand{\tr}{\mbox{tr}}
\newcommand{\slsh}[1]{/ \!\!\!\! #1}

\newcommand{\1}{\mbox{1}\hspace{-0.25em}\mbox{l}}

\def\vac{|0\rangle}
\def\lvac{\langle 0|}

\def\hlf{\frac{1}{2}}
\def\ove#1{\frac{1}{#1}}
\newcommand{\hot}[1]{\frac{#1}{2}}

\def\Box{\square}
\def\CC {\mathbb{C}}
\def\FF {\mathbb{F}}
\def\RR{\mathbb{R}}
\def\NN{\mathbb{N}}
\def\ZZ{\mathbb{Z}}
\def\bb#1{{\bf #1}}
\def\bcomment#1{}
\def\bfhat#1{{\bf \hat{#1}}}
\def\VEV#1{\left\langle #1\right\rangle}

\newcommand{\ex}[1]{{\rm e}^{#1}} \def\ii{{\rm i}}

\newcommand{\lrbrk}[1]{\left(#1\right)}
\newcommand{\lrsbrk}[1]{\left[#1\right]}
\newcommand{\sfrac}[2]{{\textstyle\frac{#1}{#2}}}

\def\stw{{\sqrt{2}}}

\def\rf {{\rm f}}
\def\ri {{\rm i}}
\def\rj {{\rm j}}
\def\rn {{\rm n}}
\def\rk {{\rm k}}
\def\rl {{\rm l}}
\def\rr {{\rm r}}

\def\rQ {{\scriptscriptstyle \rm \cQ}}
\def\rR {{\scriptscriptstyle \rm \cR}}

\def\cQb{{\cal \Qb}}
\def\cRb{{\cal \Rb}}
\def\cWb{{\cal \Wb}}

\def\fd {{\rm N}}
\def\afd {{\overline{\rm N}}}

\def \II {I\hspace{-.1em}I\hspace{.1em}}
\def \IIA {\mbox{\II A\hspace{.2em}}}
\def \IIB {\mbox{\II B\hspace{.2em}}}
\def \gs {g^s}
\def \ls {\lambda^s}

\def \I {{\cal I}}
\def \qs {q\hspace{-.53em}/\hspace{.15em}}
\def \ks {k\hspace{-.53em}/\hspace{.15em}}
\def \YM {{\mbox{\tiny YM}}}
\def \gym {g_{\YM}}

\def \Lc {\L_c}
\def\IR{\relax{\rm I\kern-.18em R}}
\def \id {{\bf 1}}

\def\cci{\ell}
\def\ccj{\ell'}

\def\bbq{\pmb{q}}
\def\bom{\pmb{\o}}
\def\bJ{\pmb{J}}
\def\bM{\pmb{M}}
\def\bB{\pmb{B}}
\def\bn{\pmb{n}}
\def\bE{\pmb{E}}

\title{\Large \bf\boldmath Holography for Boundary Lifshitz Field Theory}

\author[a,b,c]{Chong-Sun Chu}
\author[a]{Ignacio Garrido Gonzalez}
\author[b,c]{Himanshu Parihar}

 \affiliation[a]{
	Department of Physics, National Tsing-Hua University,
 Hsinchu 30013, Taiwan}
\affiliation[b]{Center of Theory and Computation,
National Tsing-Hua University,
 Hsinchu 30013, Taiwan}
\affiliation[c]{Physics Division,
    National Center for Theoretical Sciences,
     Taipei 10617, Taiwan}

\emailAdd{cschu@phys.nthu.edu.tw}
\emailAdd{ignacio.garrido@gmail.com}
\emailAdd{himansp@phys.ncts.ntu.edu.tw}

\abstract{\noindent We propose a holographic duality for the boundary
  Lifshitz field theory (BLFT). 
  Similar to holographic BCFT, holographic BLFT
  can be consistently defined
  by imposing either a Neumann boundary condition (NBC) or a
  conformal boundary condition (CBC) on the end of the world (EOW) brane.
  We propose $g$-functions and derive $g$-theorem for these
  two types of holographic BLFT. On the field theory side, we consider BLFT
  whose path integral is prescribed to include
  also paths bouncing off the boundary. The entanglement entropy for an interval
  for the Lifshitz invariant
  ground state is computed in the saddle point approximation,
  and is found to agree precisely with the holographic
  result in both limits when the interval is very close or very far away from
  the boundary. }

\begin{document}
	
	\maketitle
	\flushbottom
	\pagebreak

	\definecolor{orange}{rgb}{1.0, 0.49, 0.0}

\section{Introduction}

One of the most fascinating aspects of quantum gravity that have come
to be recognized is that quantum spacetime appears to be holographic
at the fundamental level \cite{tHooft:1993dmi,Susskind:1994vu}. This
has found realization in the celebrated duality of AdS/CFT
\cite{Maldacena:1997re}. Not only that, it has also been recognized
that holography appears to be entropic and quantum
informatic in nature \cite{Ryu:2006bv,Ryu:2006ef},
which appears to be a fundamental aspect of quantum space time. The original
AdS/CFT duality was formulated with a conformal symmetry in the field
theory.
However it is not only
not needed in the formulation of holography
\cite{tHooft:1993dmi,Susskind:1994vu}, but in fact a duality without
conformal symmetry is even richer in physics since it allows to
capture the RG flow \cite{deHaro:2000vlm} of the quantum field theory.  In the
studies of holographic renormalization, there is a full conformal
symmetry at the fixed point of the RG flow.  From this point of view,
it is interesting to consider holography of QFT with less conformal
symmetry at the fixed point. One such possibility is to consider a
boundary QFT.  In \cite{Takayanagi:2011zk,Fujita:2011fp}, a
holographic duality of AdS/BCFT was proposed for a boundary conformal
field theory (BCFT) where the gravity dual is given by a portion of
the AdS spacetime bounded by an
end of the world (EOW) brane. The
location of the EOW brane is determined by a boundary condition of the bulk
gravity, which can be a Neumann boundary condition (NBC) as originally
proposed in \cite{Takayanagi:2011zk,Fujita:2011fp}, or a conformal
boundary condition \cite{Miao:2017gyt,Chu:2017aab} or a Dirichlet
boundary condition \cite{Miao:2018qkc} as later discovered. All these
boundary conditions (BC) give rise to a consistent duality of
AdS/BCFT.

In  this regard, it is interesting to consider a theory with Lifshitz
scaling symmetry of the form
\be
t \to \l^z t, \quad x^i \to \l x^i, \quad \l >1
\ee and consider its holography \cite{Kachru:2008yh,
  Taylor:2008tg,Taylor:2015glc}.  Such holographic description is
particularly useful since it provides window to study quantum critical
point and RG flow of non-relativistic field theoretic
systems. Previously, field theoretic realization of Lifshitz symmetry
has been known for special integer value of the dynamical exponent
$z=2$ as quantum Lifshitz model (QLM) \cite {Ardonne_2004} and various
bipartite entanglement measures were subsequently analyzed within QLM
in \cite{Fradkin:2006mb,Fradkin:2009dus,
  PhysRevB.80.184421,Hsu:2008af,Hsu:2010ag,Oshikawa:2010kv,
  PhysRevLett.107.020402,Zhou:2016ykv,MohammadiMozaffar:2017nri,
  Berthiere:2019lks,Boudreault:2021pgj,Angel-Ramelli:2019nji,
  MohammadiMozaffar:2017chk,Angel-Ramelli:2020wfo,10.21468/SciPostPhys.12.4.134,Berthiere:2023bwn}. The
QLM was later generalized to $(d+1)$-dimensions with $z=d$ by
demonstrating exact matching of correlation functions in both field
theory and bulk descriptions \cite{Ker_nen_2012, Keranen:2016ija,
  Park:2022mxj}.
Recently the Lifshitz scalar field theory with
arbitrary dynamical exponent $z$ was constructed in
\cite{Basak:2023otu} (see also \cite{Benedetti:2023pbt,Benedetti:2024oif}). It was found that the Lifshitz scale invariant
ground state of the theory takes on the special form of
Rokhsar-Kivelson (RK) \cite{PhysRevLett.61.2376, Henley_2004}, and
because of this, many entanglement quantities can be computed
explicitly at ease even without the aid of twist operators as in the
case of CFT  \cite {Calabrese:2004eu,Calabrese:2009qy}. Moreover, the entanglement entropy obtained was found to
be in perfect agreement with the holographic result, giving support to
the Lifshitz holography and the Ryu-Takayanagi formula \cite{Ryu:2006bv,Ryu:2006ef} in this less
symmetric setting.

In this paper, we are interested in boundary Lifshitz field theory (BLFT) and
its holography. This duality can be expected to be useful for studying the
boundary effects in strongly coupled non-relativistic systems and
quantum critical points. In \cref{Hol-BLFT},  we
follow the construction of \cite{Takayanagi:2011zk,Fujita:2011fp} and propose a
holography dual for BLFT by imposing a boundary condition
on the so called EOW brane $Q$. 
The nontrivial part is that the 
bulk Lifshitz gravity background has now also in presence a
massive gauge field whose boundary condition is
found to be
generally violated.
We find  quite remarkably that a certain
boundary gauge field action term can be
added such that the modified boundary conditions for the gauge field
as well as the metric can be satisfied.
We show that this work for both the Neumann boundary
condition (NBC) as well as conformal boundary condition (CBC) on the metric.
In \cref{g-theorem}, we consider the $g$-theorem \cite{Affleck:1991tk} for boundary RG flow in BLFT.
The holographic $c$-theorem for LFT has been constructed in \cite{Chu:2019uoh}.
Here we establish a $g$-theorem for holographic BLFT defined by both the
NBC and the CBC. Interestingly, the proposed $g$-functions are different for
the two types of holographic BLFT and their monotonicity are guaranteed by
two different energy theorems imposed on the boundary matters on the EOW brane.
In \cref{EE-BLFT}, we consider the field theory side.
We define a BLFT by giving a consistent prescription
of the path integral that incorporates the contributions of paths that bounced
off the boundary. Using this, we can compute the entanglement entropy
of an interval for the Lifshitz ground state of the BLFT. We show that
the result agrees exactly with the holographic result in the limit when the
interval is either very close or very far away from the boundary. Finally
in \cref{summary} we conclude with the summary and discussions.

\section{Proposal for holographic BLFT}\label{Hol-BLFT}

We begin by considering the metric for $(d+1)$-dimensional Lifshitz spacetime
with one-direction anisotropy as
\begin{equation}
  ds^2=L^2\left[-\frac{dt^2}{r^{2z}}+\frac{dr^2}{r^2}
    +\frac{d \vec{x}^2}{r^2}\right],
 \label{poincare metric}
\end{equation}
where $L$ is the length scale and $\vec{x}$ denotes the $(d-1)$ spatial
directions. The above metric is invariant under the non-relativistic
Lifshitz scaling
transformation
\begin{equation}
  t\rightarrow\lambda^{z}t, \quad \vec{x} \rightarrow \lambda \vec{x} ,
  \quad r\rightarrow
  \lambda r,
	\label{L-scaling}
\end{equation}
which is consistent with the Lifshitz symmetry of the dual field theory.
The metric (\ref{poincare metric}) together with the gauge field configuration
\begin{equation}\label{A-Lifshitz}
  A\equiv A_M dx^M =\sqrt{\frac{2(z-1)}{z}}\frac{L}{r^z}dt
\end{equation}
appears as a solution to the equation
of motion of the following AdS gravity action with  a massive gauge field,
\begin{equation}
  S=\frac{1}{16\pi G_N}\int d^{d+1}x\sqrt{-g}
  \left(R- 2 \L -\frac{F^2}{4}-\frac{1}{2}M^2A^2\right).
      \label{Lif-bulk-action}
\end{equation}
The metric \eq{poincare metric} and the gauge field \eq{A-Lifshitz}
solves the equation of motion if the mass is fixed to be a specific value
\be
M^2=\frac{(d-1)z}{L^2}
\ee
and $L$ is related to the cosmological constant as
\be \label{Lambda}
\L = - \frac{z^2+(d-2)z+(d-1)^2}{2L^2}: = -\frac{f(z)}{2 L^2}.
\ee
Note that $f=d(d-1)$ when $z=1$ as expected.
We remark that in \cite{Taylor:2008tg}, the cosmological constant is fixed
as $\Lambda =-d(d-1)/2$ which fixes the length scale 
$L$ to be some complicated function of $z$ and $d$.
However, it is more convenient to allow $L$ to be a free
parameter for discussing holography
\cite{Kachru:2008yh}.
The reality of fields and
positivity constraints the value of $z\geq 1$, so we will consider
this case here.

We now proceed to construct the holographic dual of the Lifshitz field
theories defined on a $d$-dimensional manifold $M$ with a boundary $\del M$.
Following the construction
of AdS/BCFT \cite{Takayanagi:2011zk,Fujita:2011fp}, let us extend $M$ to a $(d+1)$-dimensional manifold
$N$ such that $\del N = M \cup Q$, where $Q$ is a $d$-dimensional manifold
called the
end of the world (EOW) brane  and
satisfies $\del Q = \del M$.
The indices of the bulk $N$ are denoted by the capital Roman letters
$L, M, N = 0, 1, \cdots, d$ and the indices of the
$d$-dimensional manifolds $M$ and $Q$ are denoted by Greek letters
$\m, \n$ etc. 
The bulk dual to a boundary Lifshitz field theory (BLFT)
is given by the gravity theory in
a portion of the Lifshitz spacetime \eq{poincare metric} bounded by $Q$. 
To make sense of the variation problem,
a Gibbons–Hawking boundary term is needed and one considers the action,
\be
  I_0=  \frac{1}{16\pi G_N}\int_N d^{d+1}x\sqrt{-g}
  \left(R - 2\L -\frac{F^2}{4}
  -\frac{1}{2}M^2A^2\right) +\frac{1}{8\pi G_N}\int_Q d^{d}x \sqrt{-h}(K-T),
      \label{I0}
\ee
where
$K$ is the scalar extrinsic curvature of $Q$, $h_{ij}$ is the induced
metric on $Q$ and $T$ is an arbitrary parameter which can be
interpreted as the tension of $Q$.
Taking the variations of (\ref{I0}), and focusing
on the boundary terms, we have
\begin{equation}\label{variation-bdy-term}
  16 \pi G_N \delta I_0= \int_Q (K^{\a\b}-(K-T)h^{\a\b}
  )\sqrt{-h}\delta h_{\a\b}
  +\int_Q \sqrt{-h}\; n_Q^M F_{M N} \delta A^N,
\end{equation}
where $n_Q$ denotes the  normal
vector on $Q$. The variation of the metric part give rises to
different type of boundary conditions on $Q$. One is the
Neumann boundary condition (NBC) 
\begin{equation}\label{nbc0}
K_{\a\b}-(K-T)h_{\a\b} = 0
\end{equation}
as originally proposed by
\cite{Takayanagi:2011zk,Fujita:2011fp}
which arises from an arbitrary variation
$\d h_{\a\b}$ of the boundary metric.
The Neumann boundary condition 
imposes conditions on $Q$ \cite{Takayanagi:2011zk,Fujita:2011fp,Nozaki:2012qd}
as well as the bulk Einstein metric \cite{Miao:2017aba}.
Apart from this, it is possible to impose a conformal boundary condition (CBC)
\cite{Miao:2017gyt,Chu:2017aab}
\be
K=\frac{d T}{d-1}, \label{cbc0}
\ee
which arises from varying 
the trace part of the boundary metric $\d h_{\a\b} = 2 \d \s h_{\a\b}$.
The conformal boundary condition fixes
the conformal geometry and the trace of the extrinsic curvature of
$Q$ \cite{Chu:2021mvq}.

In addition, we have also from
the gauge field variation part in (\ref{variation-bdy-term})
the boundary condition 
\be\label{A-bc}
n_Q^M F_{MN}\Pi^N_{\ \a} =0,
\ee
where  $\Pi$ is the projection operator
which gives the vector field and metric on $Q$:
$a_\b=A_M \Pi^M_{\ \b}$, $h_{\a\b}= g_{MN}\Pi^M_{\ \a} \Pi^N_{\ \b}$. 
The gauge field boundary condition \eq{A-bc} in AdS/BCFT
has been analyzed in the past \cite{Chu:2018ntx,Chu:2018fpx,Chu:2019rod}
with interesting results. In 4-dimensions, 
the boundary gauge field \cite{Chu:2018ntx,Chu:2018fpx}
has led to the phenomena of magnetically induced quantum current in holographic
BCFT. This was originally discovered as a fundamental effect arising
from the Weyl anomaly in BCFT \cite{Chu:2018ksb}. The holographic
analysis of the boundary gauge field has been extended to 6-dimensional BCFT
in \cite{Chu:2018fpx,Chu:2019rod} and has led to the prediction of an
$N^3$ degrees of freedom in
the non-Abelian 2-form gauge theory of $N$ M5-branes for $N$ large. 
We note that in all these cases, the boundary condition \eq{A-bc} is satisfied
by a configuration of field strength with the nonzero components
$F_{ra}, F_{xa}$, where
 we have chosen to parameterized
$M$ with the Gauss normal coordinates $x^\mu = (t, x, y^a)$, with
$x \geq 0$ describing the position away from the boundary $\del M$
and $y^a$ refers to the orthogonal coordinates to $x$.

In our present gauge field configuration \eq{A-Lifshitz}, 
we have only a nonzero $F_{r0}$ and obviously the boundary condition
\eq{A-bc} cannot be satisfied for any  $Q$ as long as $n^r \neq 0$.
To overcome this problem, we propose to modify the BC (\ref{A-bc}) by
adding a boundary
action term  on $Q$: 
\begin{equation} \label{I1}
  16 \pi G_N I_1= \int_Q \sqrt{-h} \;  \; \frac{1}{2}\mu M^2 a^2
  \; \cos \th ,
\end{equation}
where $a_\m$ is the projected gauge field on $Q$,
$\th$ is the angle between the normal of $Q$ and the normal of $M$
and $\mu$ is a constant of length dimension given by
\be
\mu = \frac{1}{d-1} \sqrt{\frac{f(z)}{2 |\L|}}
\ee
with $ f(z) = z^2 + (d-2)z + (d-1)^2$ as defined in \eq{Lambda}. 
All in all, we propose that the holographic dual of BLFT is
described by the bulk
gravity-gauge action given by $I = I_0 + I_1$ of \eq{I0} and \eq{I1}. 
Including the variation from $I_1$,
the BC (\ref{A-bc}) is modified to
\begin{equation}\label{m-A-bc}
n_Q^N F_{N \a} + \mu M^2 \cos \th \,  a_\a=0,
\end{equation}
and the NBC and the CBC are modified to 
\bea
{\rm NBC:} && K_{\a\b}-(K-T)h_{\a\b}=T^A_{\a\b}, \label{nbc1}\\
{\rm CBC:}  && d \,T-(d-1)K=T^A, \label{cbc1}
\eea
where
\begin{equation}\label{TA}
  T^A_{\a\b}=\frac{1}{2}\mu M^2 \cos \th
  \left(a_\a a_\b-\frac{1}{2}h_{\a\b} a^2\right),
  \quad T^A=\left(1-\frac{d}{2}\right) \frac{1}{2}\mu M^2 \cos \th \; a^2
\end{equation}
are boundary energy-momentum tensor
obtained from the gauge action term $I_1$,
$16 \pi G_N\delta I_1 :=\int_Q \sqrt{-h}T_A^{\a\b} \delta h_{\a\b}$.
For the usual AdS/BCFT, $M=0$ and there is no need of the boundary gauge field
$a_\a$ and so the boundary conditions \eq{m-A-bc}, \eq{nbc1}, \eq{cbc1}
reduce back to the usual ones \eq{A-bc}, \eq{nbc0}, \eq{cbc0}.
In general with a massive vector field, the holographic BLFT with $z>1$ is
defined by the modified BCs
\eq{m-A-bc}, \eq{nbc1}-\eq{TA}.

For explicitness, let us analyze the case
of 2-dimensional BLFT 
defined on the half space $x\geq 0$. 
For the present massive background,
we have $\mu M^2 = z/L \neq 0$ and 
the vector field $a_\a$ has the non-vanishing component,
\begin{equation} \label{a_0}
a_0=\sqrt{\frac{2(z-1)}{z}}\frac{L}{r^z}.
\end{equation}
It is easy to check that the BC \eq{m-A-bc} is satisfied for the
components $\a \neq 0$.   For the component $\a =0$,
let us consider an embedding of $Q$ described
by $x =x(r)$. 
This gives the unit normal vector
\be \label{nQ}
(n_Q^t, n_Q^x, n_Q^r) = \frac{r}{L}\frac{1}{ \sqrt{1+ x'^2}}(0,1, -x')
\ee
and hence
$\cos \th = \frac{ -x'}{ \sqrt{1+ x'^2}}$.
It follows
immediately that \eq{m-A-bc} is satisfied for $\a \neq 0$.
Next, let us consider the metric BC \eq{nbc1} and \eq{cbc1}. 
The extrinsic curvature $K_{\a\b} = \frac{1}{2} \cL_n g_{MN} e^M_\a e^N_\b$
can be obtained from the normal vector. We obtain 
\be
K_{rr} = \frac{L}{r^2 \sqrt{1+x'^2}} \left(-r x'' + x'(1+x'^2)\right),
\qquad
K_{tt} = -\frac{L z x'}{ r^{2z}  \sqrt{1+x'^2}}.
\ee
Using  the induced metric on $Q$
\be \label{ind-metric}
ds^2 = L^2\left(-\frac{dt^2}{r^{2z}} + (1+x'^2)\frac{dr^2}{r^2}\right),
\ee
we obtain the trace as
\begin{equation}\label{trace-K}
  K=h^{\a\b}K_{\a\b}=\frac{1}{L}\left(\frac{(z+1)x'}{(1+x'^2)^{1/2}}
  -\frac{x'' r}{(1+x'^2)^{3/2}}\right).
\end{equation}
The case of two-dimensional BLFT is interesting since $T^A=0$.
As a result,  the form of CBC
\begin{equation} \label{CBC2}
K=2T
\end{equation}
is not modified.
It then follows immediately that the CBC \eq{CBC2} is satisfied with
\be \label{xr}
x = r \sinh \left(\frac{\rho_*}{L}\right),
\ee
if $T$ is parameterized in terms of $\rho_*$ as
\be \label{Trho}
T =\frac{z+1}{2L}\tanh \left(\frac{\rho_*}{L}\right). 
\ee
The expression \eq{Trho} generalizes the tension of the EOW brane
in AdS/BCFT \cite{Takayanagi:2011zk,Fujita:2011fp} to general $z \neq 1$.
As before $\rho_*$ admits a geometric meaning since we can perform a
coordinate transformation from $(t,r,x)$ to the Gauss-normal
coordinates $(t,\rho,y)$ as
\begin{equation}
  r=\frac{y}{\cosh \left(\rho/L\right)}, \hspace{4mm}
  x=y \tanh \left(\frac{\rho}{L}\right).
\end{equation}
The metric \eq{poincare metric} becomes
\be
  ds^2=d\rho^2+L^2\left(\cosh^2\left(\frac{\rho}{L}\right)
  \frac{dy^2}{y^2}-\frac{1}{y^{2z}}\cosh^{2z}
  \left(\frac{\rho}{L}\right)dt^2\right).
  \ee
This gives 
\begin{equation}\label{K-EOW}
K=\frac{z+1}{L}\tanh \left(\frac{\rho}{L}\right),
\end{equation}
showing immediately
that the CBC is  solved if the EOW brane is located at a constant
$\rho=\rho_*$ and with $\rho_*$ determined from $T$ through \eq{Trho}.
Finally, let us consider the NBC. Unlike the CBC, the NBC is given
by a tensor
multiplet of equations on the embedding $x=x(r)$
and it
is not clear if it admit a solution in general.
Nevertheless, for the present Lifshitz background, one can check
that \eq{xr} solves the modified NBC non-trivially.
To see this, note that
the gauge field configuration \eq{a_0} gives rise to $T^A_{\a\b}$ of the form
$(T^A_{tt}, T^A_{rr}) = B (h_{tt}, -h_{rr})$, 
where $B:= \frac{z-1}{2L} \frac{x'}{\sqrt{1+x'^2}}$
and $h_{\m\n}$ is the induced metric \eq{ind-metric} on $Q$.
For brane profile with $x'' =0$, we have
$(K_{tt}, K_{rr}) = \frac{x'}{L \sqrt{1+x'^2}} (h_{tt}, h_{rr})$,
and it is now easy to see that the NBC is satisfied with the brane profile
\eq{xr}. It is remarkable that the same brane profile \eq{xr} universally
defines the 2d BLFT with a flat boundary, and is independent of the
non-trivial
anisotropic
modifications present
in the bulk metric \eq{poincare metric}
as well as the holographic BC \eq{nbc1}, \eq{cbc1}.

\section{Holographic $g$-theorem for 2-dimensional BLFT}\label{g-theorem}

For two-dimensional relativistic QFTs, the Zamolodchikov $c$-theorem
\cite{Zamolodchikov:1986gt} asserts
the existence of a real
function $c$ that decreases monotonically along the RG flow, with
value given by the central charge of the CFT at the RG fixed point.
See
\cite{Freedman:1999gp,Casini:2004bw,Myers:2010tj} for holographic
$c$-theorem in relativistic CFT.
For anisotropic scale invariant field
theory (this includes the Lifshitz field theories), a holographic
$c$-theorem was considered in \cite{Chu:2019uoh} where the $c$-function was
constructed in terms of the holographic entanglement entropy.
We are interested in the $g$-theorem for holographic BLFT here. 

Generally given a BCFT, one can consider a perturbation to 
a BCFT at the UV by a set of relevant
local operators
$\{ \phi_i \}$
at the boundary \cite{Casini:2016fgb},
\be
S = S_{BCFT_{UV}} + \int_{\del M} \l_i \phi_i.
\ee
The boundary perturbation triggers a boundary RG flow
which ends up at a BCFT${}_{IR}$ in the IR.
The $g$-theorem then asserts the existence of a $g$-function
which decreases monotonically along the boundary RG flow, with value
given by the boundary central charge of the BCFT at the RG fixed
point.  The $g$-theorem was originally conjectured in
\cite{Affleck:1991tk}, and was later proven in \cite{Friedan:2003yc}.
It has also been proven using the idea of 
relative entropy 
\cite{Casini:2016fgb,Casini:2022bsu}, and most recently
in \cite{Harper:2024aku}
by using the 
strong subadditivity property of entanglement entropy. 
For holographic BCFT, a version of the $g$-function was proposed in
\cite{Takayanagi:2011zk}.  There the boundary perturbation was
modeled
by a perturbation of the energy-momentum tensor
$T^Q_{\m\n}$ on the EOW brane $Q$, and the $g$-theorem was shown to
follow from
the imposition of the null energy condition on $Q$.
The construction works for
AdS/BCFT defined by NBC, but is unknown for AdS/BCFT defined by CBC
because there does not seem to be any
obvious way to take 
the null energy condition into account in the CBC.

In this paper, we follow the construction of
\cite{Takayanagi:2011zk,Fujita:2011fp} to propose a
holographic $g$-function that depends on
the brane embedding $x(r)$
such that
\begin{equation} \label{gt1}
\frac{d}{d r}\log g(r) \leq 0
\end{equation}
when an energy condition is imposed.
It is also required that
\be \label{gt2}
\lim_{r\to 0} \log g(r) = \frac{\rho_{UV}}{4G_N},
\quad
\lim_{r\to \infty} \log g(r) = \frac{\rho_{IR}}{4G_N}.
\ee
To model the boundary RG flow in BLFT, we keep the bulk metric
\eq{poincare metric} unchanged and 
consider a general distribution of matter $T^Q_{\a\b}$ on $Q$.
The boundary condition reads respectively,
\begin{align}
&{\rm NBC}: \quad K_{\a\b}-K h_{\a\b}=T^Q_{\a\b},\label{nbc2}\\
&{\rm CBC}: \quad (1-d)K=T^Q. \label{cbc2}
\end{align}
We will in fact propose two versions of $g$-function. The first
one is a direct generalization of the one proposed by \cite{Takayanagi:2011zk}
and works for holographic-BLFT
defined by NBC. The second $g$-function makes use of
a different energy condition, i.e. the dominant energy condition. This
$g$-function is non-local in $r$ (energies) but it works for holographic-BLFT
defined by both NBC and CBC. The non-locality of the $g$-function is in perfect
agreement with the fact that the monotonicity of the $g$-function is due to
an integrating-out of the high energy degrees of freedom in the theory.

\subsection{$g$-theorem for holographic BLFT with NEC}

Let us start with the analysis of the case of NBC. Consider
the null energy condition (NEC) on $T^Q_{\a\b}$:
$T^Q_{\a \b} \cN^\a \cN^\b\geq 0$, for any null vector
$N^\a$ on $Q$. 
Then \eq{nbc2} gives the geometric condition 
\begin{equation}\label{Null_EC}
K_{\a\b}\cN^\a \cN^\b-Kh_{\a\b}\cN^\a \cN^\b\geq 0.
\end{equation} 
For an embedding of brane profile $x=x(r)$, let us consider the null vector
$ (\cN^t, \cN^r) =(\pm r^{z-1},\frac{1}{\sqrt{1+x'^2}} )$,
Applying the null energy condition on the NBC implies that
\begin{equation}
\frac{(1-z)x'(1+x'^2)-rx''}{r^{2}(1+x'^2)^{3/2}}\geq 0.
\end{equation}
This allows one to conclude that the function $f:= rx' -x$ is non-positive
since $f' = rx'' \leq 0$ from the energy condition ($z\geq 1$) and that
$f(0) =0$. As a
result, we propose following definition of $g$-function
\cite{Takayanagi:2011zk}
for holographic-BLFT defined by NBC,
\begin{equation} \label{g1}
\log g(r): =\frac{L}{4G_N}\sinh^{-1}\left(\frac{x(r)}{r}\right).
\end{equation}
It is monotonically decreasing since up to a positive factor,
$ \frac{d}{d r}\log g(r) \propto f(r)\leq 0$. At the UV and IR, the
boundary perturbation modeled by $T^Q_{\a\b}$ is supposed to have been
turned off and so the brane profile is given by \eq{xr}
with $\rho_*$  given by $\rho_{UV}$ (resp. $\rho_{IR}$)
at the UV (resp. IR). As a result, \eq{g1} gives
\be \label{fixed-g}
\log g (x_{UV} (r))  = \frac{\rho_{UV}}{4G_N}, \quad
\log g (x_{IR} (r))  = \frac{\rho_{IR}}{4G_N}
\ee
and the condition \eq{gt2} is satisfied.

We remark that the construction of holographic $g$-theorem depends crucially on
an energy condition. For holographic BCFT defined by NBC, the null
energy condition (NEC) allows one to translate the NBC \eq{nbc2} to an
geometric condition on the embedding of the surface $Q$, which then
leads to a $g$-theorem for an appropriately constructed
$g$-function. This however does not work for BCFT or BLFT with
CBC \eq{cbc2} since the null energy
condition does not give rises to a condition on
the trace of the energy momentum tensor. To see this, let us consider, without
loss of generality, a diagonal metric on $Q$.  The
trace $T^Q$ is given by $T^Q =A+B$, where we have denoted
$A=T^Q_{tt}h^{tt}$ and $B=T^Q_{rr}h^{rr}$ for convenience.
On the other hand, the  null energy condition gives $A \leq B$ since
$T^Q_{\a \b}N^\a N^\b = \frac{(N^t)^2}{-h_{tt}} (-A+B)$
where $N^\a$ is  an arbitrary null vector. Obviously the null energy condition
does not constraint the trace $T^Q$.

\subsection{$g$-theorem for holographic  BLFT with DEC}

The missing of a $g$-theorem for holographic BCFT defined by CBC has been
a puzzle to the authors of \cite{Miao:2017gyt,Chu:2017aab}. In this subsection, we propose to consider the
dominant energy condition (DEC) and prove the $g$-theorem for holographic
BLFT defined by both NBC and CBC. 
The dominant energy condition (DEC) states that for every
future-pointing causal vector field $Y^\b $ (means $Y^0 >0$ and either
timelike $Y^2 <0$ or null $Y^2 =0$), the vector $u^\b =
-T^{\a}_{\b}Y^{\b}$ must be a future-pointing causal vector.
Physically, the DEC means that the local energy density
measured by any observer is always
non-negative and
that it cannot flow faster than light.
Note that since the inner
product of two timelike vectors is always negative,
the DEC implies the weak energy condition (WEC):
for every timelike vector field $W^\m$, the local energy density
$T_{\m\n} W^\m W^\n \geq 0$ is always non-negative \cite{Hawking:1973uf}.

Now consider any future pointing causal vector $Y^\a$:
\begin{equation}
h_{tt}(Y^t)^2+h_{rr}(Y^r)^2\leq 0, \quad Y^t>0.
\end{equation}
Requiring the vector $u^\a=-T^\a_\b Y^\b=(-AY^t,-BY^r)$ to be 
future pointing implies that  $A<0$.  Since
$ u^\a u_\a=A^2\left((Y^t)^2 h_{tt} +(Y^r)^2h_{rr}\right)+(B^2-A^2)(Y^r)^2h_{rr}$,
requiring $u^\a$ to be causal for any $Y^\a$ implies that $B^2 \leq A^2$.
This means  $-|A|\leq B\leq |A|$ and so 
\begin{equation}
T^Q=A+B \leq 0.
\end{equation}
The dominant energy condition for CBC thus implies that
$K \geq 0$:
\begin{equation}\label{K-dominant}
K=\frac{1}{\sqrt{1+x'^2}}\left((z+1)x'-r \frac{x''}{1+x'^2}\right) \geq 0.
\end{equation}
This condition also holds in holographic BCFT and BLFT
defined by NBC
since \eq{cbc2} follows from taking the trace of the NBC \eq{nbc2}.

Let us now consider a holographic $g$-function of the form
\be \label{g2}
\log g(r) = \frac{L}{4G_N} \left(
(c_\infty -c_0) H(r) + c_0
\right),
\ee
where
$c_\infty := \r_{IR}/L$ and 
$c_0 := \r_{UV
}/L$.
This verifies the $g$-theorem \eq{gt1}, \eq{gt2}
if the function $H$ satisfies $H(0) =0,
H(\infty) =1$ and $dH/dr \geq 0$ as a result of the energy condition
\eq{K-dominant}. Such a $H$ can be constructed easily, for example,
\be
H(r) = \int_0^r d\rt K f(\rt) /\int_0^\infty d\rt K f(\rt),
\ee
where $f$ is a positive function chosen such that the
integral in the
denominator is finite. Note that for large $r$, $x' = O(1/r)$ and so
$K = {\rm constant} + O(1/r)$, it is therefore sufficient to choose
$f = e^{-r}$ or $f = 1/(1+r^2)$. We note that the $g$-function \eq{g2}
involves an integral from the UV ($r=0)$
down to the energy scale marked by $r$. This is consistent with the
usual field theory expectation that in field theory,
the $g$-function (or the $c$-function
in CFT) is obtained by integrating out the high energy degrees of freedom.

We remark that the $g$-function \eq{g2}
works for holographic BLFT defined by both NBC or CBC, while the
$g$-function \eq{g1} only work for BLFT defined by NBC.
As a result, in a holographic BLFT defined by NBC, in addition
to the local form \eq{g1} of  the $g$-function, there is also a second
$g$-function \eq{g2}. This is not surprising as generally there can be
many $g$-function in a theory. Recall
in the original construction \cite{Zamolodchikov:1986gt},
the $c$-function was constructed from the energy-momentum tensor.  In
\cite{Casini:2004bw}, an entropic $c$-function was discovered and it
was shown to
be different from the one of Zamolodchikov.
In our present case, the
two $g$-functions appears to be physically independent since they  arise
from different energy conditions for matter on the boundary,
which would be natural for them to give rise
to different boundary perturbations, and hence  different
boundary RG flows and  different associated $g$-functions.

\section{Entanglement Entropy for BLFT}\label{EE-BLFT}

We now proceed to
examine the boundary effect on entanglement entropy in BLFT.
As an example, we
compute the entanglement entropy for an interval in
a two-dimensional BLFT. In this context, we first obtain the
holographic entanglement entropy using the RT formula which involves
the computation of extremal curves (geodesics) in the bulk dual of a
BLFT. Subsequently, we compute the entanglement entropy
on the field
theory side using the propagator of the BLFT and show that it matches
exactly with holographic results.

In a general QFT, the entanglement entropy of a region $A$ is given by
\be
S(A) = - \Tr (\r_A \log \r_A),
\ee
where $\r_A$ is the reduced density matrix obtained by tracing out the degrees
of freedom in the complementary region of $A$. One of the popular ways to
facilitate the calculation of the entanglement entropy is to consider the
R\'{e}nyi entropy defined by
\be
S_n(A) = \frac{1}{1-n} \log \Tr (\r_A^n),
\ee
where $n$ is an integer-valued R\'{e}nyi index. The entanglement entropy can be
obtained by analytically continuing
$n$ into a continuous variable and take the limit $n \to 1$
\be
S(A)= \lim_{n \to 1} S_n(A).
\ee
The value of $\Tr(\r_A^n)$ can be calculated by considering a
path integral 
on a replicated manifold $M_n$. Here the replicated manifold contains
$n$ copies of the original manifold $M$ where the QFT is defined, sewn
together cyclically along the subregion $A$ following the cyclicity of
the trace. As a result, the trace $\Tr(\r_A^n)$ is given by \cite{Calabrese:2004eu,Calabrese:2009qy}
\be
\Tr(\r_A^n) = \frac{Z_n}{Z^n},
\ee
where $Z_n = Z[M_n]$ and $Z= Z_1$ are the partition function over $M_n$ and
$M$ with marked points given by the endpoints of the interval $A$.
As a result, the entanglement entropy can be obtained through
the free energy $F_n=\log Z_n$ as 
\be
S(A)= \lim_{n \to 1} \frac{1}{1-n} (F_n - n F_1) .
\ee

\subsection{Holographic result}
We are interested in the
$(1+1)$-dimensional Lifshitz scalar field theory on a half line
$x\geq 0$ with a boundary at $x=0$
\be \label{blft}
L=\frac{1}{2}\int_0^\infty dx \left[\left(\partial_{t}\phi\right)^2
  -\kappa^2\left(\del_x^z \phi\right)^2\right].
\ee
In general, the definition of fractional
derivative is not unique and can be chosen to suit the application in
mind \cite{Herrmann:2011zza}.
For field theory, it is useful to adopt
a definition of the fractional derivative in terms of the plane wave modes
\be
\del_x^z e^{ikx} = (ik)^z e^{ikx}.
\ee
As described in \cite{Basak:2023otu} for the full Lifshitz theory without boundary, the
field theory vacuum for general $z >1$ is given by the RK vacuum, whose 
properties are quite distinct from the usual conformal vacuum of a
conformal field theory. 
This can be seen, for instance, in the properties of the entanglement
entropy. For example, the entanglement entropy for an interval
$A$ of length $l$ in an infinite system is given by \cite{Basak:2023otu}
\be \label{SA}
S(A) = \frac{z-1}{2 }\log\frac{l}{\epsilon}. 
\ee
Note that the factor $(z-1)$ appears instead of the central charge $c$ for
a CFT. In \cite{Basak:2023otu}, it was found that
the field theory result \eq{SA} can be reproduced from holography
if the Lifshitz metric curvature scale $L$ is related to the 3-dimensional
Newton constant via
\be
\label{LG}
L=(z-1)G_N.
\ee
This relation is
appropriate for the RK vacuum of the Lifshitz theory
and is different from the Brown-Henneaux relation $c=3L/2G$
for the CFT vacuum \cite{Brown:1986nw}.

\begin{figure}[H]
			\centering
                        \includegraphics[scale=1.5]{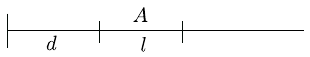}
			\caption{Schematic of a single interval on a half line.}
			\label{Single-interval-fig}
\end{figure}

Obviously the result \eq{SA} will have to be modified in the presence of
boundary, especially when $A$ is near the boundary.
For an interval $A$ of length $l$ at a distance $d$ from the boundary
(see  \cref{Single-interval-fig}), 
the entanglement entropy can be obtained
immediately  on the holographic side. 
In fact, as we discussed above, 
the holographic BLFT with NBC or CBC are well defined and
is given by the dual gauge-gravity action in a portion of the Lifshitz Poincar\'{e}
metric bounded by $Q$: 
\begin{equation}\label{Poincare-metric-1}
\begin{aligned}
ds^2&=L^2\left[-\frac{dt^2}{r^{2z}}+\frac{dr^2}{r^2}+\frac{dx^2}{r^2}\right],\\
Q:& \hspace{2mm} x=r \sinh\left(\frac{\rho_*}{L}\right).
\end{aligned}
\end{equation}
The tension of the EOW brane lies in the range $0\leq T \leq \frac{z+1}{2L}$.
For the interval $A$ at a constant time,  the metric  (\ref{Poincare-metric-1})
at constant time slice is identical to that of the
AdS$_3$/BCFT$_2$.
Using the Ryu-Takayanagai formula \cite{Ryu:2006bv}
\begin{equation}
S(A)=\frac{\ell_A}{4 G_N},
\end{equation}
where $\ell_A$ is the length of the geodesic in the bulk which is
homologous to the boundary subsystem $A$,
the holographic entanglement entropy is the same as that given by
AdS/BCFT \cite{Chu:2017aab}:
\begin{numcases}{S(A)=}
\frac{L}{2 G_N}\log\frac{l}{\epsilon},      \hspace{4cm} d \geq d_c  \\
\frac{\rho_*}{2 G_N}+\frac{L}{4G_N}\log\frac{4d(d+l)}{\epsilon^2},
\quad\quad \quad\,\, d \leq d_c,
\end{numcases}
where the critical distance $d_c$ is determined when the above expressions
cross over
\footnote{
The emergence of a new saddle point in the holographic entanglement entropy
and the interplay between this and the original RT surface
was originally noticed
in \cite{Chu:2017aab}. Explanation in terms of BCFT channels was  observed
in \cite{Sully:2020pza}.
}.
Note however the curvature radius $L$ is related to the
Newton constant via a
different relation. Replacing $L$ with $G_N$ using \eq{LG}, we obtain
the holographic entanglement entropy for a single interval in BLFT as
\begin{numcases}
{S(A) = }
\frac{z-1}{2 }\log\frac{l}{\epsilon},  \qquad
\qquad d\geq d_c \label{HEE-a} \\
\frac{z-1}{4}\log\frac{2d}{\epsilon}
+\frac{z-1}{4}\log\frac{2(d+l)}{\epsilon}+\frac{\rho_*}{2G_N},
\quad\quad \quad\,\, d\leq d_c.\label{HEE-b} 
\end{numcases}

When the interval is far away from the boundary $d \gg l$,
one can neglect the effects of boundary and the result  agrees
with \eq{SA}.
The other limit 
where the interval is near to the boundary $d \ll l$ is more interesting.
In this case,  we have the holographic entanglement entropy (\ref{HEE-b}).
Our goal is to perform the computation
on the field theory side
and check it
against \eq{HEE-b}. This requires an understanding of the effect of boundary
on the path integral computation of entanglement entropy for BLFT. Before
we analyze this,
let us recall the field theory computation for the entanglement entropy in LFT.

\subsection{Path integral and the entanglement entropy in Lifshitz field
theory}

The computation of $Z_n$ is generally quite complicated. In a CFT,
the computation is greatly simplified due to the availability of
the twist operator in the replica formalism \cite{Calabrese:2004eu,Calabrese:2009qy}.
There is however no such thing in LFT. Fortunately,
the computation of $Z_n$ is also very simple
due to the special form of the vacuum in a Lifshitz field
theory.
Recall that 
\cite{Basak:2023otu} the ground state in Lifshitz scalar field theory
takes on the special form of RK 
\begin{equation} \label{RK}
  |\Psi_0\rangle=\frac{1}{\sqrt{\mathcal{Z}}} \int \cD \phi \;
  e^{-S_{\rm cl}[\phi]/2} |\phi\ra, \quad
   S_{\rm cl}[\phi] :=  \kappa
  \int \left(\nabla_x^{\frac{z}{2}}\phi\right)^2 dx ,
\end{equation}
where
\begin{equation}
\mathcal{Z}=\int \mathcal{D}\phi e^{-S_{\rm cl}[\phi]}
\end{equation}
is  a normalization factor. As a result,
the partition function $Z_n$ is given \cite{Basak:2023otu} by an integral over
the products of the path integral propagator 
\be \label{K-LFT}
  K(x_1, \phi_1;x_2, \phi_2) = \int_{\phi(x_1)
= \phi_1}^{\phi(x_2) =\phi_2}
  \cD \phi
  e^{-S_{\rm cl}[\phi]}.
  \ee
  In the saddle point approximation, the propagator is determined by the
  solution to the equation of motion. For the transition from
  $\phi(x_1) =\phi_1$ to   $\phi(x_2) =\phi_2$, the solution is 
 not unique. This give rises
  to a family of ground states and the kernel $K$ is given by
\be \label{K-free}
K_{\rm free}(x_1, \phi_1; x_2, \phi_2) = \sqrt{\frac{\g}{\pi (x_2-x_1)^{z-1}}}
e^{-\g\frac{ (\phi_2-\phi_1)^2}{(x_2-x_1)^{z-1}}}.
\ee
Here $\g$ is a constant which parameterizes the family of solution
\cite{Basak:2023otu}. For example, for the
classical solution
\begin{equation}
\phi=\phi_1+(\phi_2-\phi_1)\frac{(x-x_1)^{z-1}}{(x_2-x_1)^{z-1}},
\end{equation}
it is 
\begin{equation}
\g=\frac{1}{z-1}\left(\frac{\Gamma(z)}{\Gamma(z/2)}\right)^2.
\end{equation} 
So far this is for LFT without boundary.
We like to extend this computation to BLFT.

\subsection{Entanglement entropy computation in BLFT}

In order to compute the entanglement entropy in BLFT,
one needs to understand how to implement boundary condition on 
the Lifshitz field theory. In a standard Lagrangian field theory, the boundary
condition can be derived from the requirement of
a well defined variational principle of the action. This leads to the
Neumann BC or Dirichlet BC on the fundamental field, which subsequently
leads to a boundary condition on the path integral kernel. To understand this, 
let us consider
the wave functional $\Psi$ of the theory. 
In terms of the kernel $K$, the wave functional 
at $\phi(x)$ can be obtained from all possible source $\phi'$ at $x'$ as
\begin{equation}
\Psi(x,\phi)=\int D \phi' K(x,\phi;x', \phi')\Psi(x', \phi').
\end{equation}
This allows one to obtain the boundary condition on $K$ from the more obvious
boundary condition on $\Psi$. 
For Dirichlet boundary condition, we have $\phi(x=0)=c$, for some constant and so
\be
\Psi(0,c)=0, \quad \forall\, c.
\ee
This translates to the boundary condition on $K$
\begin{equation}\label{K-DBC}
  {\rm DBC}: \quad K(x, \phi ;x', \phi')\Big |_{x=0} =0,
  \quad \forall \,\,\phi, \phi',x'.
\end{equation} 
For Neumann boundary condition, we have
$\partial_x \phi(0)=0$ and the wavefunctional satisfies
\begin{equation}
\partial_x\Psi(x,\phi)\Big |_{x=0}=0.
\end{equation}
This translates to the following boundary condition on $K$
\begin{equation}\label{K-NBC}
  {\rm NBC}: \quad \partial_x
  K(x, \phi;x', \phi' )\Big |_{x=0}=0, \quad \forall \,\,\phi, \phi',x'.
\end{equation}
Such $K$ for the boundary theory can be constructed 
with the inclusion of images. See the
appendix \ref{K-QM} for a
brief review for the construction of propagator in boundary QM
and boundary QFT.

\begin{figure}[H]
			\centering
                        \includegraphics[scale=0.9]{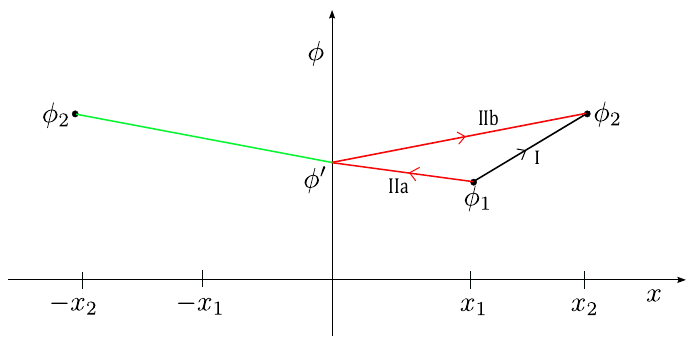}
			\caption{Schematic of paths I and II used for
                          computation of the propagator.}
			\label{K-BLFT-fig}
		\end{figure}

For the Lifshitz field theory \eq{blft} defined on the half space,
it is not possible to extract the boundary conditions on the field
from the action principle since  there isn't an integration by part
formula for the fractional derivative $\del_x^z$.
On the other hand, for the sake of doing QFT
computation, it is sufficient to be able to take into account of the
effect of boundary into the path integral through \eq{K-DBC} or \eq{K-NBC}.
It is interesting to note that the holographic result (\ref{HEE-b}) is
given by half of the total length of the geodesic which extends from
end points of the interval and intersects at the EOW brane while going
to their corresponding mirror image \cite{Chu:2017aab}. This suggests that
images must continue to serve important roles in the field theory side
of BLFT. Let us consider the definition \eq{K-LFT} for the propagator
in the presence of boundary. As shown in  \cref{K-BLFT-fig}, there are two
saddle points for the transition from
\be
\label{tt}
\phi(x_1) = \phi_1 \quad\mbox{to}\quad \phi(x_2) = \phi_2.
\ee
For the path I,
the propagator for this path is given by \eq{K-free}.
Now in presence of a boundary, one may also have the
path II as a new saddle point.
Denote the point of reflection $\phi'$, then the
classical actions for the path IIa and the path IIb are given respectively
by
$  S_{\rm cl, IIa}=\int_0^{x_1}(\partial_x^z\phi)^2 dx$, 
$S_{\rm cl, IIb}=\int_0^{x_2}(\partial_x^z\phi)^2
dx=\int_{-x_2}^{0}(\partial_x^z\phi)^2 dx$,
and the classical action can be evaluated as
$S_{\rm cl, II}=\int_{-x_2}^{x_1}(\partial_x^z\phi)^2 dx$ for the path
$\phi=\phi_1+(\phi_2-\phi_1)\left(\frac{x+x_2}{x_1+x_2}\right)^{z-1}$.
This gives
$S_{\rm cl,II} =\gamma
\frac{(\phi_2-\phi_1)^2}{(x_2+x_1)^{z-1}}$.
As a result, we can construct the propagator for the transition \eq{tt}
by summing the contributions from the two saddle points. We obtain
\footnote{
We remark that in the literature, another class of boundary condition
on the Lifshitz field theory has been considered.  Since we are
interested in $K(x_i, \phi_i; x_f, \phi_f)$, if we compare it with the
$K(t_i,x_i; t_f, x_f)$ of QM, it suggests a definition of $K$ by
replacing $x\to \phi$, $t\to x$. This gives the following kernel
\begin{equation} \label{K-half}
  K(x_1, \phi_1; x_2,\phi_2)
  = e^{-\g
  (\phi_2-\phi_1)^2/(x_2-x_1)^{z-1}}\mp e^{-\g
  (\phi_2+\phi_1)^2/(x_2-x_1)^{z-1}},
\end{equation}
which has been considered in \cite{Chen:2017txi,Chen:2017tij,
  Boudreault:2021pgj} in the context of continuum version of the
ground state for
the Motzkin and Fredkin spin chains. However this is
not the desired propagator for the boundary theory since \eq{K-half}
satisfies the conditions $K |_{\phi=0}=0$ or $\partial_\phi K
|_{\phi=0}=0$. These are not boundary conditions in the $x$-space, but
instead the field space admits a boundary $\phi \geq 0$.  This results
in a so called positive Lifshitz theory. It is not clear if such a
theory is consistent quantum mechanically. In any case, it is
different from the boundary Lifshitz theory we considered here.
}
\begin{equation}\label{K-BLFT}
K(x_1,\phi_1; x_2,\phi_2) = \sqrt{\frac{\g}{\pi (x_2-x_1)^{z-1}}}
e^{-\g \frac{(\phi_2-\phi_1)^2}{(x_2-x_1)^{z-1}}}\mp
\sqrt{\frac{\g}{\pi (x_2+x_1)^{z-1}}} e^{-\g
  \frac{(\phi_2-\phi_1)^2}{(x_2+x_1)^{z-1}}}.
\end{equation}
It is interesting to see that  \eq{K-BLFT}
satisfies the boundary conditions \eq{K-DBC} and
\eq{K-NBC} with the $-$ and $+$ choice of sign respectively.
Although we cannot derive the conditions  \eq{K-DBC}, \eq{K-NBC} 
from a fundamental action principle, we can utilize a formulation of the
quantum theory by specifying a consistent definition of the path integral.
This trick is widely practiced in string theory, for example,
the path integral for multiple M5-brane \cite{Witten:1996hc} is
formulated without a worldvolume theory.

\begin{figure}[H]
			\centering
                        \includegraphics[scale=0.9]{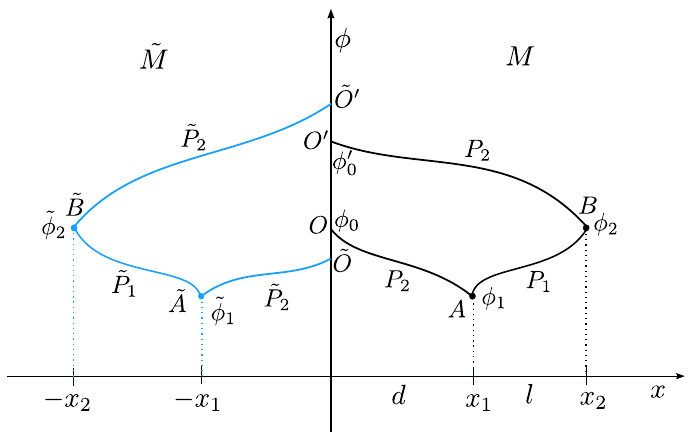}
			\caption{Schematic of possible paths for the
                          propagator in a BLFT.}
			\label{K-entropy-single-fig}
		\end{figure}

For us, let us for simplicity consider the case with the $+$ choice
of sign and  define a version of boundary
Lifshitz field theory by declaring that its path integral is given by
a sum over all possible direct paths and bounced paths, i.e. those that
involve a discontinuity at the boundary.
For example, the propagator $K$ for
transition from a point  $A = (x_1,\phi_1)$ to a point $B = (x_2, \phi_2)$
is given by
\be
K(A;B) = \int_{\{P_1\}} \cD \phi e^{-{S_{\rm cl}}}
+  \int_{\{P_2\}} \cD \phi e^{-{S_{\rm cl}}} ,
\ee
where $P_1 /P_2$ are arbitrary
direct/bounced paths in $M$ that connect $A$ to $B$.
In the saddle point approximation, it reduces to \eq{K-BLFT}.
In figure
\ref{K-entropy-single-fig}, we show an example of a bounced path $P_2$ 
that involves an absorption
at the boundary at $\phi_0$ and an emission off the boundary
at a generally different value $\phi_0'$ of the field. The classical solution
path II in figure \ref{K-BLFT-fig} corresponds to a special configuration of
$P_2$ with special value of $\phi_0 =\phi_0'$.

Now, we can follow the calculation as for the LFT and obtain
\be
Z_n =
\int d\phi_1 d\phi_2 d\phi_0 d\phi_0'
\;  K(O,A)^n K(A,B)^n K(B,O')^n,
\ee
where the points in field space are $O= (0,\phi_0)$,  $O'= (0,\phi_0')$,
$A =(x_1,\phi_1)$ and $B= (x_2, \phi_2)$ as shown in figure \ref{K-entropy-single-fig}.
Schematically, we have
$Z_n = \int_{P= P_1, P_2} e^{-nS(P)}$, where the integration over path is a shorthand
for the path integral.
To facilitate the computation of $Z_n$, it is useful to consider a copy of
the boundary theory on the negative side $x \leq 0$ with a similar set of
paths $\Pt_1, \Pt_2$ and end points $\Ot= (0,\phit_0)$,  $\Ot'= (0,\phit_0')$,
$\At =(x_1,\phit_1)$ and $\Bt= (x_2, \phit_2)$.
Consider
the product $Z_n^2 = (Z_n)_{x\geq 0} (Z_n)_{x \leq 0}$. We have
\be\label{Zn2}
Z_n^2 = I_1+ I_2+I_3
\ee
where
\be \label{I12}
I_1:= \int_{P_1,\Pt_1} e^{-nS(P_1)} e^{-nS(\Pt_1)},\qquad
I_2: = \int_{P_2,\Pt_2}  e^{-nS(P_2)} e^{-nS(\Pt_2)},
\ee
and
\be \label{I3}
I_3:= \int_{P_1,\Pt_2}  e^{-nS(P_1)} e^{-nS(\Pt_2)} +
\int_{\Pt_1,P_2}  e^{-nS(\Pt_1)} e^{-nS(P_2)} .
\ee
Note that the term $I_2$ can be written as a sum of the path connecting 
$A \At$ and $B \Bt$, including discontinuous ones.
In the
saddle point approximation, the expression \eq{Zn2}
is dominated by the solution to the equation of motion.
For general configurations of $d, l$, all three terms are important. 
It is easy to see that in the far region where
$d \gg l$, the calculation simplifies and the
term $I_1$ dominates in \eq{Zn2}.
Since the first term in \eq{K-BLFT} dominates in this regime, we obtain
in this limit, 
\be
Z_n = \int D\phi_1 D\phi_2 \left(\frac{\g}{\pi l^{z-1}}\right)^{n/2}
e^{-n\g (\phi_2-\phi_1)^2/l^{z-1}}
={\rm  const.} \times \frac{1}{\sqrt{n}}(\frac{\g}{\pi l^{z-1}})^{\frac{n-1}{2}}
\ee
This gives
the R\'{e}nyi entropy and the entanglement entropy as
\begin{equation}
S_n(A) =\frac{(z-1)}{2}\log \frac{l}{\epsilon} + {\rm const.}, \qquad  
S(A) =\frac{(z-1)}{2}\log \frac{l}{\epsilon} + {\rm const.},
\end{equation}
where the constants are $z$-independent.
The leading logarithmic term
agrees precisely with the corresponding holographic
computation (\ref{HEE-a}).

The path integral \eq{Zn2} also simplifies in the
near region $d \ll l$ where the term $I_2$ dominates.
In the saddle approximation, the path integral $I_2$ is given by classical
paths from $A$ to $\At$, and $B$ to $\Bt$.
Using the propagator which connects $-x_1$ to $x_1$, 
\begin{equation}\label{K-III}
K(-x_1,\phit_1; x_1, \phi_1)= \sqrt{\frac{\g}{\pi (2x_1)^{z-1}}}
e^{-\g (\phit_1-\phi_1)^2/(2x_1)^{z-1}},
\end{equation}
and for $-x_2$ to $x_2$, 
\begin{equation}\label{K-IV}
K(-x_2, \phit_2; x_2, \phi_2)= \sqrt{\frac{\g}{\pi (2x_2)^{z-1}}}
e^{-\g (\phit_2-\phi_2)^2/(2x_2)^{z-1}}.
\end{equation}
We obtain
\be
\begin{aligned}
  Z^2_n&=
  \int D\phi_1 D\phi_2 D\phit_1 D\phit_2
  \left(\frac{\g}{\pi (2d)^{z-1}}\right)^{n/2}
  e^{-\frac{n\g (\phit_1-\phi_1)^2}{(2d)^{z-1}}}
\left(\frac{\g}{\pi (2(d+l))^{z-1}}\right)^{n/2}
e^{-\frac{n\g (\phit_2-\phi_2)^2}{(2(d+l))^{z-1}}}\\
&
={\rm const.}\times\frac{1}{n}(\frac{\g}{\pi})^{n-1}
\left(\frac{1}{(2d)^{z-1} \; (2(d+l))^{z-1}}\right)^{\frac{n-1}{2}}.
\end{aligned}
\end{equation}
This gives the R\'{e}nyi entropy and the entanglement entropy as
\be  \label{sn-field}
S_n(A) =\frac{z-1}{4}\log\frac{2d}{\epsilon}
+\frac{z-1}{4}\log\frac{2(d+l)}{\epsilon}+{\rm const.},
\ee
\be\label{s-field}
S(A) =\frac{z-1}{4}\log\frac{2d}{\epsilon}
+\frac{z-1}{4}\log\frac{2(d+l)}{\epsilon}+{\rm const.}
\ee
We observe that \eq{s-field} agrees with the  corresponding
bulk result (\ref{HEE-b}) for the leading logarithmic terms in the near
boundary limit.
As in 2d BCFT, the
boundary entropy appears in the subleading orders to the logarithmic
divergence, we expect it to hold similarly in BLFTs.
It will be interesting
to compute the boundary entropy by going beyond the leading
saddle point approximation.
We  note that our result for
the R\'{e}nyi entropy for an interval of length $2l$ in a LFT
without boundary and the R\'{e}nyi entropy 
for an interval $[0,l]$ adjacent to the boundary in a BLFT
obey the relation
\be \label{SSS}
S_n (2l) = 2 S_n^{(B)}(l)
\ee
for the leading logarithmic term.
The relation  \eq{SSS},  which relates the universal terms in R\'{e}nyi
entropies for certain theories with and without a boundary,
was proposed originally in \cite{Berthiere:2019lks} and it has been
verified for different classes of QFTs
including LFTs in $(2+1)$ dimensions with $z=2$.

\section{Summary and discussion}\label{summary}

To summarize, we have studied the Lifshitz field theory in the presence of
a boundary and constructed its holographic dual that is given by a
portion of Lifshitz spacetime truncated by the
end of the world brane. The holographic bulk dual is constructed by
adding a boundary gauge field action term in addition to the usual
Gibbons–Hawking boundary term,
and satisfies the modified boundary
conditions for both the gauge field and metric. We found that for
BLFT defined on a half space,
NBC and CBC leads to the same brane profile regardless of the
anisotropic nature of the bulk metric. We then introduced two types of
holographic $g$-functions, each relying on a different
energy condition on the EOW brane,
namely the null energy condition (NEC) and the dominant
energy condition (DEC). The first one make use of null energy
condition (NEC) with NBC and was shown to decrease monotonically along
the boundary RG flow. However, since NEC doesn't impose constraints on
the 
trace of the
energy-momentum tensor, a
holographic $g$-theorem  for CBC that is based on the NEC was not
possible. To address this, we proposed a second $g$-function utilizing
the dominant energy condition (DEC) and showed that it works for both
NBC and CBC. Thus we
established  the general validity of holographic $g$-theorems in BLFT.

Subsequently, we computed the entanglement entropy for an interval
holographically from the bulk dual to a BLFT. This involves the
computation of RT surface in a portion of the
Lifshitz Poincar\'{e}
metric. The holographic entanglement entropy is determined by,
depending on the interval's distance from the boundary, two
choices of the RT surface. Following this,
we described how to implement suitable
boundary conditions in Lifshitz field theory and obtained the
propagator in a BLFT. The propagator has contributions from the paths
which are bounced off the boundary in addition to the usual path of a
free theory. When the interval is far away from the boundary, the free
path gives the dominant contribution in the saddle point approximation
of the propagator. Conversely, when the interval is very close to the
boundary, the reflected path dominates in the path integral of the
propagator. Then we computed the entanglement entropy for an interval
in a BLFT by employing the propagator in both the near-boundary and
far-boundary limits. Interestingly, we observed that the entanglement
entropy for an interval in these two limits matches
precisely with the
corresponding holographic entanglement entropy results.

There are many interesting directions to follow from our work in this
paper. It would be
interesting to construct an entropic $g$-function
for a BLFT using the methods from quantum information theory similar
to \cite{Casini:2016fgb,Casini:2022bsu,Harper:2024aku}. It would also
be interesting to try to generalize our approach to defect RG flows
and higher dimensions. In this work, we focused on computing the
entanglement entropy for BLFT at zero temperature case, however
extending our analysis to non-zero temperature in the context of
boundary Lifshitz theory still remains a non-trivial issue. Similar to
the approach in \cite{Chu:2023zah}, it would be interesting to
investigate the phase structure of timelike entanglement entropy in
BLFT.  Finally, it would also be interesting to explore the
codimension two or wedge holography \cite{Akal:2020wfl} as a generalization of the
Lifshitz holography. We leave these open issues for future
investigation.

\section*{Acknowledgments}
C.S.C thanks Pei-Ming Ho for helpful discussion.
C.S.C acknowledge support of this work by NCTS and the grant
113-2112-M-007-039-MY3 of the National
Science and Technology Council of Taiwan. H.P acknowledges the support
of this work by NCTS.

\begin{appendices}
  \section{$K$ in Boundary QM and Boundary QFT}\label{K-QM}
  
\begin{figure}[H]
  \centering
  \includegraphics[scale=0.9]{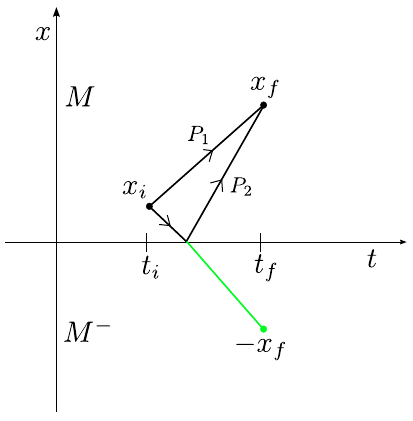}
\caption{Schematic of particle trajectories in the presence of a boundary.}
			\label{K-QM-fig}
		\end{figure}

For quantum mechanics (QM) without boundary, it is sufficient to define
the theory by a path
integral which integrates over all
possible paths as
\begin{equation} \label{Z-QM}
  Z=\int Dx(t) e^{-S}.
\end{equation}
However, this prescription is not sufficient in the presence of a
boundary since we also need to know how a field is reflected/absorbed
from a boundary in order to include the correct path in the path integral.
Suppose the wavefunction is given by $\psi(t_1,x)$ at time $t_1$.
Then the wave function at a later time $t_2$ is given by
\be
\psi(t_2,x) = \int dy K(t_1,y;t_2,x) \psi(t_1,y),
\ee
where
\be
K(t_1,x_1; t_2,x_2 )=\int_{x(t_1)=x_1}^{x(t_2)=x_2}
Dx(t) e^{-S}
\ee
and the integration measure has to be defined so that $K$ satisfies
some boundary condition.
The boundary condition on $K$ follows from the boundary condition imposed on
the wave function. For QM on a half line $x\geq 0$, the Dirichlet BC reads
$\psi(t,0) =0$ and the Neumann BC reads $\del \psi (t,x)|_{x=0} =0$. This
translates to the following BC on 
the path integral propagator:
\begin{align}
&{\rm Dirichlet \,BC}: \quad K(t_1,y;t_2,x)=0, \quad {\rm for}\,\, x=0,\,\,
  y \geq 0,\\ &{\rm Neumann \,BC}: \quad \partial_x K(t_1,y;t_2 ,x)=0, \quad {\rm
    for}\,\, x=0,\,\, y \geq 0.
\end{align}
In the saddle point approximation,
the propagator for the boundary QM is then given by \cite{Bastianelli:2006hq}
\begin{equation} \label{KK}
K(0,y;T,x)=\frac{1}{\sqrt{2\pi T}}e^{-|x-y|^2/2T}\mp
\frac{1}{\sqrt{2\pi T}}e^{-|x+y|^2/2T},
\end{equation}
where $-$ and $+$ signs are for the Dirichlet BC and Neumann BC respectively.
Note that 
the second term in \eq{KK} is due to
the reflected path $P_2$ as shown in the figure \ref{K-QM-fig}.

The discussion for QFT is similar.
Consider a QFT defined on a $d$-dimensional
spacetime with $x^\mu=(t,\vec{x})$ and
consider a state $\ket \Psi$ of a QFT. In the basis
$\ket{\phi(\vec{x}),t}$, the state $\ket \Psi$ give rise to a
wavefunctional as
\begin{equation}
\bra{\phi(\vec{x}),t}\Psi\rangle=\Psi(t,\phi(\vec{x})).
\end{equation}
Let us consider the evolution of the wavefunction from time $t_i$ to
$t_f$ as
\begin{equation}
  \bra{\phi_f(\vec{x}),t_f}\Psi\rangle
  =\int\bra{\phi_f(\vec{x}),t_f}\phi_i(\vec{x}),t_i\rangle
  \bra{\phi_i(\vec{x}),t_i}\Psi\rangle
  D\phi_i(\vec{x}),
\end{equation}
where we have inserted identity operator $\int \ket{\phi_i(\vec{x})}
\ \bra {\phi_i(\vec{x})}\,\, D\phi_i(\vec{x})=\id$ at $t_i$. Introduce
the path integral propagator $K$
\begin{equation}
  K(t_i,\phi_i(\vec{x});t_f,\phi_f(\vec{x}))
  :=\bra{\phi_f(\vec{x}),t_f}\phi_i(\vec{x}),t_i\rangle
  =\int_{\phi(t_i,\vec{x})=\phi_i(x)}^{\phi(t_f,\vec{x})=\phi_f(x)}
  e^{-S[\phi]} D\phi,
\end{equation}
then
\begin{equation}
\Psi(t_f,\phi_f(\vec{x}))=\int
D\phi_i(\vec{x})K(t_i,\phi_i(\vec{x});t_f,\phi_f(\vec{x}))
\Psi(t_i,\phi_i(\vec{x})).
\end{equation}
Now consider the half space
$\vec{x}=(x,\vec{y}, x\geq 0)$, we have the following BC on $\ket
\Psi$ as
\begin{align}
&{\rm Dirichlet \,BC}: \quad\ket \Psi=0, \quad {\rm at}\,\,
  x=0,\\ &{\rm Neumann \,BC}: \quad \partial_x\ket \Psi=0, \quad {\rm
    at}\,\, x=0,
\end{align}
which leads to
\begin{align}
&{\rm DBC}: \quad \Psi(t,\phi(x,\vec{y}))|_{x=0}=0, \quad {\rm for
    \,\,any}\,\, \phi(0,\vec{y}),\\ &{\rm NBC}: \quad
  \partial_x\Psi(t,\phi(x,\vec{y}))|_{x=0}=0, \quad {\rm for \,\,any}
  \,\, \phi(x,\vec{y}), \,\,{\rm such \,\, that}\,\, \partial_x\phi
  |_{x=0}=0.
\end{align}
This translates to the following boundary conditions on the
propagator:
\begin{align}
&{\rm DBC}: \quad K(t_i,\phi_i(\vec{x});t_f,\phi_f(\vec{x}))|_{x=0}=0, \quad
  {\rm for\,\, any}\,\, \phi_f(0,\vec{y})=0,\\ &{\rm NBC}: \quad
  \partial_x K(t_i,\phi_i(\vec{x});t_f,\phi_f(\vec{x}))|_{x=0}=0, \quad {\rm
    for\,\, any}\,\, \phi_f(x,\vec{y}),\,\, {\rm such \,\, that}\,\,
  \partial_x\phi_f|_{x=0}=0.
\end{align}

\end{appendices}

\bibliographystyle{JHEP}

\bibliography{Bdy-LFT}  
  
\end{document}